\documentclass[letter,aps,twocolumn]{revtex4-1}
\usepackage{graphics}
\usepackage{graphicx}
\usepackage{amssymb}
\usepackage{epsfig}
\usepackage{color}
\usepackage{ulem}
\newlength{\upit}\upit=0.1truein

\newcommand{\ltappr}{{{\lower4pt\hbox{$<$} } \atop \widetilde{ \ \ \ }}}
\newlength{\bxwidth}\bxwidth=1.5 truein

\newcommand{\bk}{{\bf{k}}}

\newlength{\figwidth}
\newlength{\shift}
\newcommand \bea {\begin{eqnarray} }
\newcommand \eea {\end{eqnarray}}

\begin{document}

\title{Giant isotropic Nernst effect in an anisotropic Kondo semimetal}

\author{Ulrike Stockert$^{1}$, Peije Sun$^{1,\ast}$, Niels Oeschler$^{1}$, Frank
Steglich$^{1}$, Toshiro Takabatake $^{2}$, Piers Coleman$^{3,4}$, Silke
Paschen$^{5}$}

\affiliation{$^1$Max Planck Institute for Chemical Physics of Solids, 01187
Dresden, Germany}
\affiliation{$^2$ADSM, Hiroshima University, Higashi-Hiroshima 739-8530, Japan}
\affiliation{$^3$Center for Materials Theory, Rutgers University, Piscataway, NJ
08855, USA} 
\affiliation{$4$ Department of Physics, Royal Holloway, University of London, Egham, Surrey TW20 0EX, UK}
\affiliation{$^5$Institute of Solid State Physics, Vienna University of
Technology, Wiedner Hauptstr.\ 8-10, 1040 Vienna, Austria}

\begin{abstract}
The ``failed Kondo insulator'' CeNiSn has long been suspected to be a nodal
metal, with a node in the hybridization matrix elements. Here we carry out a
series of Nernst effect experiments to delineate whether the severely
anisotropic magnetotransport coefficients do indeed derive from a nodal metal or
can simply be explained by a highly anisotropic Fermi surface. Our experiments
reveal that despite an almost 20-fold anisotropy in the Hall conductivity, the
large Nernst signal is isotropic. Taken in conjunction with the magnetotransport
anisotropy, these results provide strong support for an isotropic Fermi surface
with a large anisotropy in quasiparticle mass derived from a nodal
hybridization. 
\end{abstract}


\maketitle

There is a wide and growing interest in electron materials with topologically
protected excitation spectra, including $Z_{2}$ topological insulators and
topological superconductors \cite{Fu07.1,Moo10.1,Ali12.1} and, most recently,
topologically protected Weyl semimetals \cite{Wan11.2,Yan11.4}. Rare earth heavy
fermion systems have recently emerged as a new venue to explore the interplay of
strong correlations with topology \cite{Dze10.1,Dez16.1,Den13.1,Xu14.1s}: the
strong electron-electron interactions and spin-orbit coupling make these systems
ideal candidates for research in this area. The class of Kondo insulators, such
as SmB$_{6}$, has received much attention as candidate strongly interacting 
$Z_{2}$ topological insulators. The little-known family of Kondo semimetals
\cite{Ike96.1,Mor00.1,Nak96.1,Nak96.2} may provide a second example of such
topological protection. These compounds are considered to be failed Kondo
insulators, in which the hybridization gap contains a node that closes the gap
in certain directions, giving rise to a semimetal with a pseudogap. Transport
studies on these materials have confirmed the presence of a large anisotropy in
the magnetotransport, but such anisotropies are not in themselves an indication
of a nodal hybridization, and could derive from anisotropic Fermi surface
geometries.

In this paper we carry out a series of magneto-thermoelectric measurements on
the Kondo semimetal CeNiSn. They reveal that unlike the Hall conductivity, which
is highly anisotropic, the large Nernst effect is essentially isotropic. We show
how this unexpected isotropy rules out an anisotropic Fermi surface geometry and
is a natural consequence of cancellations between mean free path and mass
anisotropies expected in a nodal semimetal. This definite understanding of the
material's bulk properties is an important prelude to any future studies of
putative surface contributions.

CeNiSn is a heavy electron material with emergent semimetallic properties. When
it develops coherence at low temperatures, a pseudogap opens in its electronic
density of states, as revealed by both tunneling \cite{Eki95.1,Dav97.1} and
nuclear magnetic resonance \cite{Kyo90.1,Nak94.1} studies. Modest magnetic
fields are sufficient to remove the pseudogap \cite{Dav97.1}. The material
exhibits marked anisotropy in its magnetotransport properties 
\cite{Tak96.2,Ina96.1,Tak98.1,Ter02.1}. To account for this unusual behavior,
Ikeda and Miyake proposed a hybridization model for this Kondo lattice system
\cite{Ike96.1}, treating it as a Kondo insulator in which the hybridization gap
contains a node along the crystallographic $a$-axis. The presence of this node
leads to a ``V''-shaped electronic density of states, a feature which is
consistent with both tunneling and NMR measurements
\cite{Eki95.1,Dav97.1,Kyo90.1,Nak94.1}. 

An important aspect of this problem which has not received much attention is the
Fermi surface and momentum space structure of CeNiSn. One of the most striking
features of CeNiSn is the anisotropy in the Hall conductivity. As will be shown
below (Fig.\,\ref{ratio}), it is almost 20-fold between orbits within the basal
($bc$-)plane and those that are perpendicular to it. Conventionally, Hall
conductivity anisotropies are associated with corresponding anisotropies in the
mean free path: according to the Ong formula, the Hall conductivity is given by
\cite{Ong91.1}
\begin{equation}\label{Ong}
\sigma_{xy} = \frac{2e^{2}}{h} \frac{B_z}{\Phi_{0}}
\int \frac{dk_{z}}{2\pi}A_{l}(k_{z})
\end{equation}
where
\begin{equation}\label{Al}
A_{l}(k_{z}) = \vec{z}\cdot\oint\frac{\vec{l}\times d\vec{l}}{2}
\end{equation}
is the area swept out by the mean free path vector
\begin{equation}\label{l}
\vec{l}(\vec{k}) = \vec{v}_\bk \tau_\bk = \frac{1}{\hbar}\nabla_{\bk}E_{\bk} \tau_\bk
\end{equation}
as $\vec{k}$ moves around the Fermi surface on orbits perpendicular to the
applied magnetic field $\vec{B} = (0, 0, B_z)$. $\vec{z}$ is the unit vector
along $z$ and $\Phi_{0}=h/e$ is the flux quantum.

A 20-fold anisotropy in the Hall conductivity thus requires a corresponding
anisotropy in the electronic mean free paths. Although this large anisotropy has
been interpreted in terms of the nodal hybridization model, {\sl a priori} the
most natural interpretation would be a one-band model with a severely
anisotropic Fermi surface. Only in combination with the isotropic Nernst effect
presented here we can eliminate this possiblilty, and provide a definitive
interpretation in terms of a nodal Kondo semimetal.

\begin{figure}[b!]
\centerline{\includegraphics[width=1.\columnwidth]{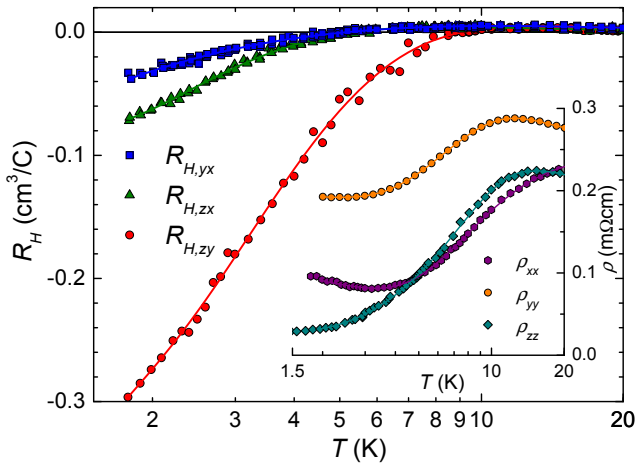}}
\caption{(Color online) Temperature dependence of the Hall coefficient $R_H$ and
of the electrical resistivity $\rho$ (inset) of CeNiSn for three different
configurations/directions. $R_H$ was measured at 1\,T, were it is in the linear
response regime. The full lines are fits to the data and serve as
guides-to-the-eyes.}
\label{Hall}
\end{figure}

High-quality single crystals grown by the Czochralski technique in a
radio-frequency furnace and purified by solid state electron transport
\cite{Nak95.1} were investigated by a steady-state heat transport technique with
one heater and differential thermocouples. To conform to the standard $x,y,z$
notation, in the following we refer to the $a$-axis as $z$, and to $b$ and $c$
as $x$ and $y$, respectively. To measure the Nernst signal $N_{yx}= E_y/\nabla_x
T$ and the Nernst coefficient $\nu_{yx} = N_{yx}/(\mu_0 H_z)$ we apply a
temperature gradient $\nabla_x T$ along $x$ and a magnetic field $\mu_0 H_z$
along $z$, which generates an electrical field $E_y$ along $y$. Other $N_{ij}$
are obtained via cyclic index permutations. The transverse temperature gradient
$\nabla_y T$ due to the Righi-Leduc effect was estimated to be less than 2\% of
the longitudinal gradient $\nabla_x T$, making any thermopower contributions to
the large Nernst signal negligible. Thus, the experimentally realized adiabatic
condition with thermally floating side edges of the sample corresponds, for our
samples, approximately to the isothermal condition that is preferentially
treated by theories. In addition, measurements of the electrical resistivity and
the Hall coefficient were performed in a commercial physical property
measurement system (Fig.\,\ref{Hall}).

\begin{figure}[tb]
\centerline{\includegraphics[width=1.8\columnwidth,angle=90]{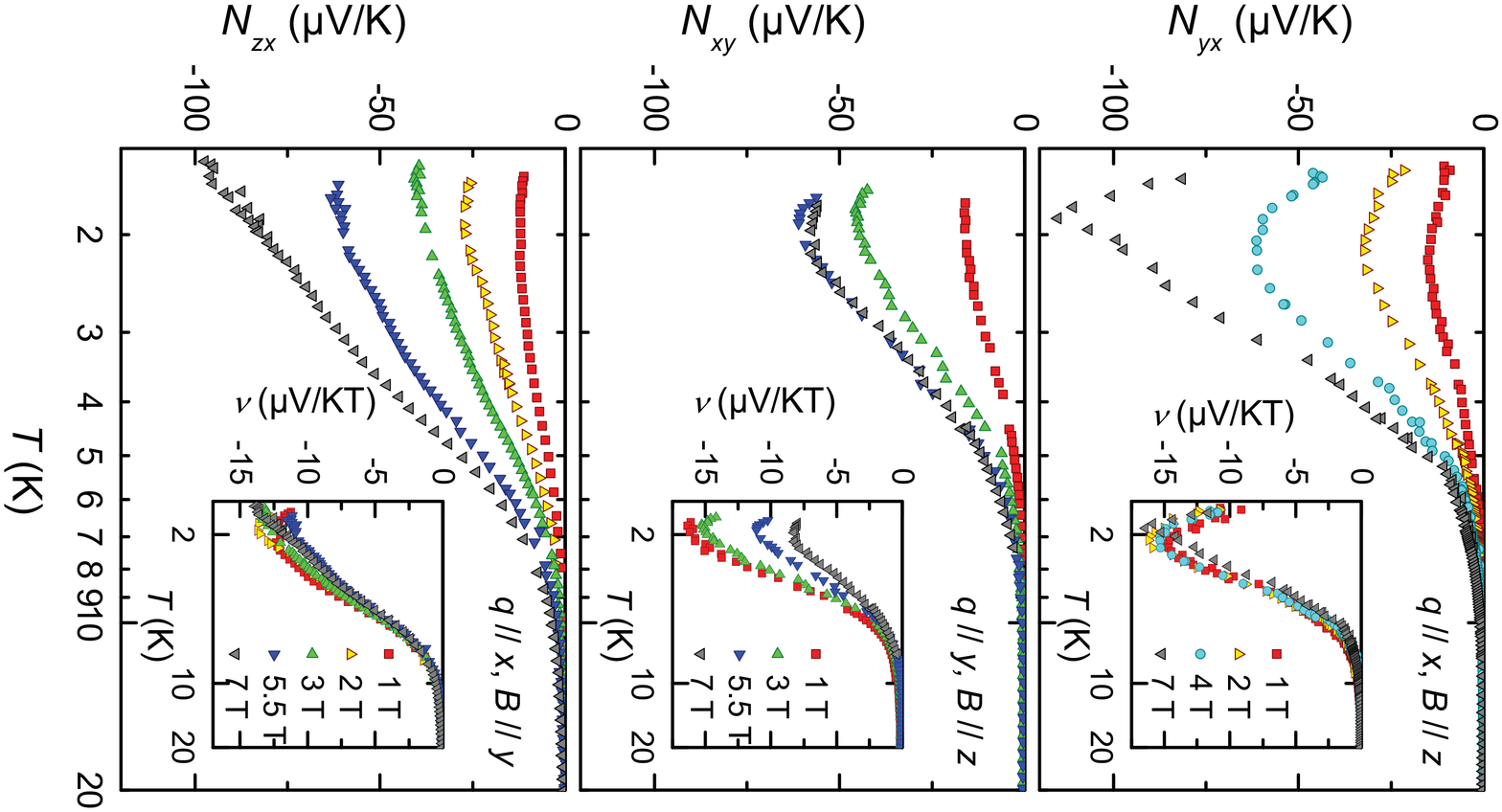}}
\vspace{-0.1cm}

\caption{(Color online) Temperature dependence of the Nernst signal $N_{ij}$ and
Nernst coefficient $\nu = \nu_{ij}$ (inset) of CeNiSn for three different
configurations, in magnetic fields up to 7\,T.}
\label{NvsT}
\end{figure}

\begin{figure}[tb]
\centerline{\includegraphics[width=1.\columnwidth]{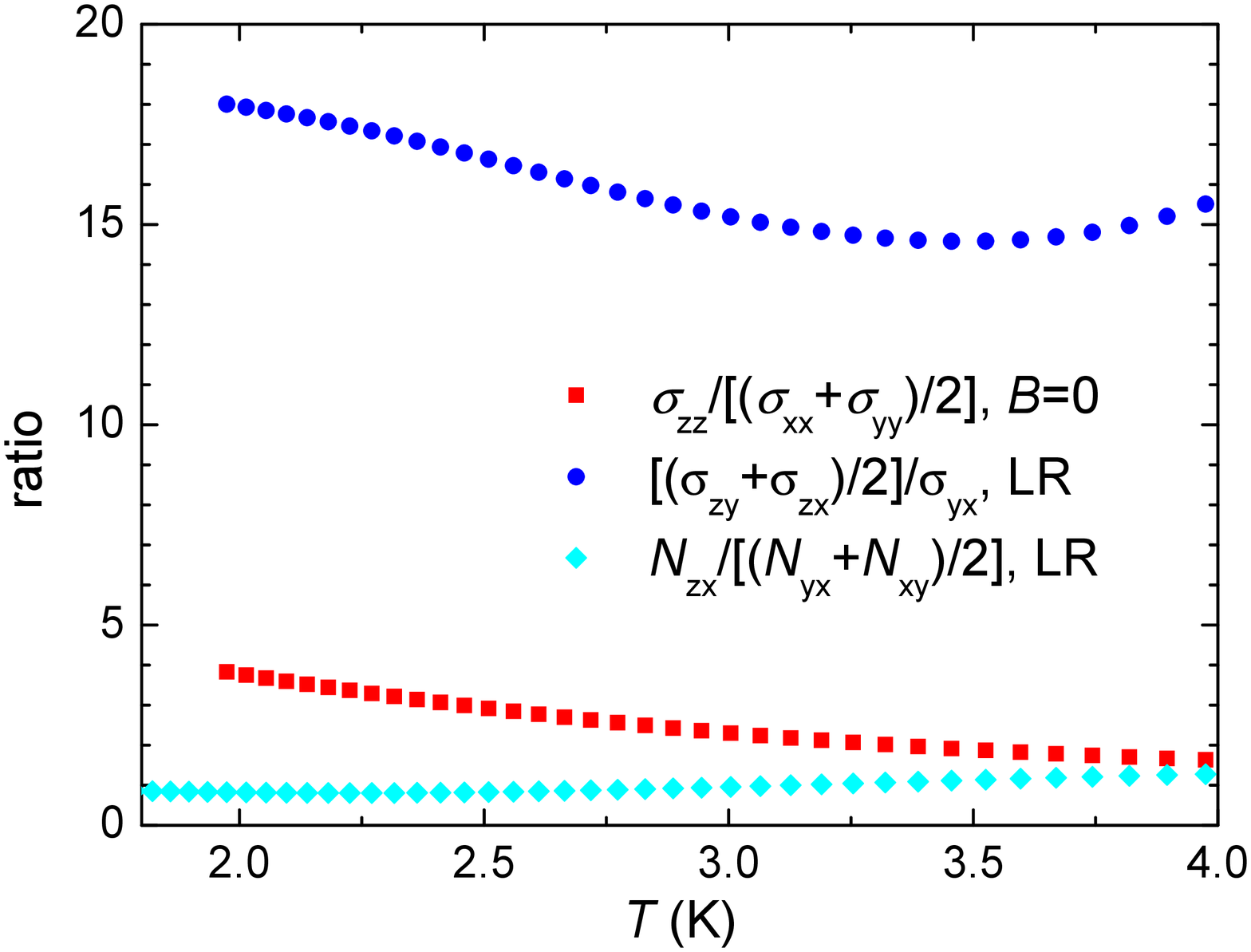}}
\caption{(Color online) Ratios of the electrical and Hall conductivity, and the
Nernst signal along the putative nodal direction $z$ to the corresponding
quantity averaged in the plane perpendicular to $z$ (see legend). Smooth fits to
the data (from Fig.\,\ref{Hall} for $\sigma_{ii}$ and $\sigma_{ij}$ and from
Fig.\,\ref{NvsT} for $N_{ij}$) were used to calculate the ratios. $\sigma_{ij}$
and $N_{ij}$ were taken in the linear response (LR) regime.} \label{ratio}
\end{figure}

Figure~\ref{NvsT} shows the temperature dependence of $N_{ij}$ and $\nu_{ij}$
(insets) for three different configurations in different magnetic fields. The
absolute values of $N_{ij}$ (and $\nu_{ij}$) strongly increase below 10\,K as
the pseudogap opens. The maximum value of 120\,$\mu$V/K reached for $|N_{yx}|$
at 7\,T and 1.8\,K is by a factor of 1.2, 4, and 130 larger than in the
previously studied $f$-electron based ``giant'' Nernst effect compounds
PrFe$_4$P$_{12}$ \cite{Pou06.1}, URu$_2$Si$_2$ \cite{Bel04.1}, and CeCoIn$_5$
\cite{Bel04.2}, respectively. At small magnetic fields $B < 3$\,T the Nernst
coefficient of CeNiSn is essentially field independent for all directions. In
the following discussion we concentrate on the 1\,T data, which are in the
linear response regime.

Our most striking observation is that this giant Nernst signal is practically
isotropic, in contrast to the highly anisotropic Hall response. It is even more
isotropic than the zero-field electrical conductivity $\sigma_{ii}$. This is
seen from Fig.~\ref{ratio} where we plot the ratio of the electrical
conductivity along the (putative) nodal direction $\sigma_{zz}$ to the
arithmetic mean of the conductivities in the plane perpendicular to $z$,
$\frac{1}{2}(\sigma_{xx}+\sigma_{yy})$, together with analoguous (linear
response) Hall and Nernst ratios
$\frac{1}{2}(\sigma_{zy}+\sigma_{zx})/\sigma_{yx}$ and
$N_{zx}/\frac{1}{2}(N_{yx}+N_{xy})$. At 2\,K, the Nernst ratio is of order
unity, whereas the conductivity and Hall conductivity ratios reach much larger
values of 4 and 18, respectively.

\begin{figure}[tb]
\centerline{\includegraphics[width=\columnwidth]{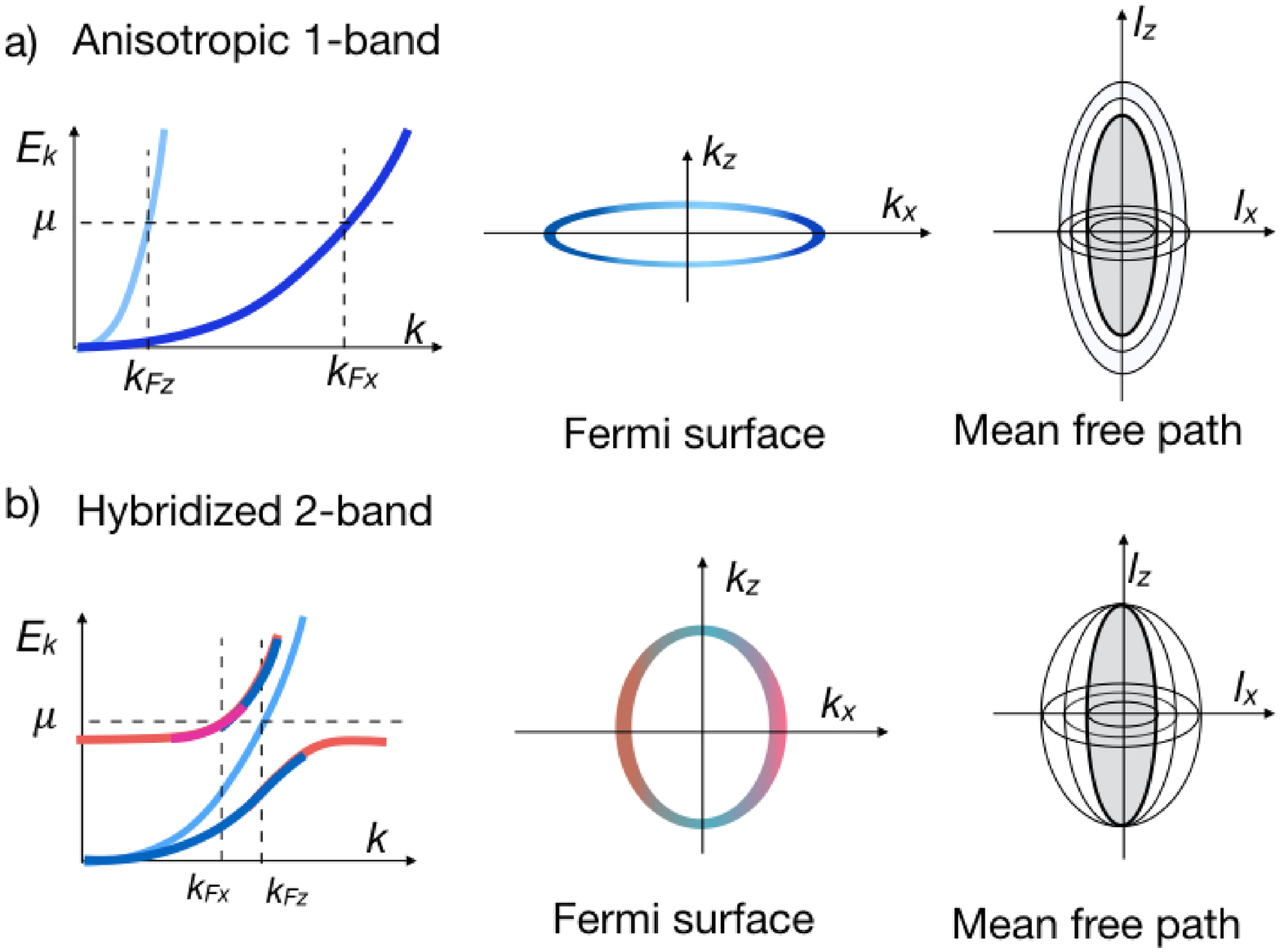}}

\caption{(Color online) Contrasting (a) one-band and (b) hybridized two-band
models of CeNiSn, showing the dispersion (left), Fermi surface (center), and
mean free path as a function of chemical potential (right). In the one-band
model, the anisotropy in the mean free path reflects the anisotropy in the
dispersion along the $z$, $x$, and $y$ directions, but in the two-band model,
the anisotropy is driven by nodes in the hybridization along the $z$-axis. Both
models give rise to mean free path anisotropy but they differ distinctly in the
dependence of this anisotropy on the position of the chemical potential. In the
one-band model the mean free path anisotropy is independent on the position of
the chemical potential, while in the two-band model, the anisotropy dependence
on chemical potential lies predominantly in the basal plane.}
\label{images}
\end{figure}

To understand this dichotomy, we now contrast a one-band transport scenario with
the nodal hybridization picture. In a one-band picture the severe Hall
anisotropy of CeNiSn, $\frac{1}{2}(\sigma_{zy}+\sigma_{zx}) \gg \sigma_{yx}$,
would conventionally be understood as a result of Fermi surface curvature. For
example, a severely flattened ellipsoidal Fermi surface (Fig.\,\ref{images}\,a)
with the dispersion ($\hbar = 1$)
\[
E_{\bk }= \frac{k_{z}^{2}}{2m} + \frac{k_{\perp }^{2}}{2m^{*}}
\]
gives rise to a reduction of order $m/m^{*}$ in the Fermi velocity in the basal
plane and, assuming an isotropic scattering rate $\tau_{\bk}$, a corresponding
reduction of the mean free path and the Hall conductivity in the basal plane. 
By contrast, in a two-band picture, strong anisotropies of the mean free path
can be driven by anisotropies in the hybridization. Suppose $V(\vec{k})\sim V
(k_{x}\pm ik_{y})$ with a node along the $z$-axis (Fig.\,\ref{images}\,b). The
corresponding dispersion is given by 
\begin{equation}
E_\bk = \frac{\epsilon_\bk+\epsilon_f}{2} \pm
\sqrt{\left(\frac{\epsilon_\bk-\epsilon_f}{2} \right)^2+ V^2k_{\perp}^2} \quad .
\label{E_hybrid}
\end{equation}
Such hybridized bands are frequently evoked to provide a simple understanding of
heavy fermion metals and Kondo insulators
\cite{Mot80.1,Aep92.1,Tsu97.1,Ris00.1}. Here $\epsilon_{\bk }={\vec{k}}^{2}/2m$
is the conduction band dispersion and $\epsilon_f$ is the position of the
$f$-level it hybridizes with. In this second scenario, the anisotropy in the
quasiparticle velocities does not derive from the Fermi surface curvature, but
from the anisotropy in the hybridization, even if the Fermi surface is
spherical. Along the $z$-axis, the quasiparticles have the conduction electron
dispersion $\epsilon_{\bk}$ with the velocity $v_{F}=k_{F}/m$, whereas within
the basal plane the hybridization with the $f$-state gives rise to a much
smaller velocity $v_{F}^{*}$, where $v_{F}^{*}/v_{F}\sim
V^{2}k_{F}^{2}/\epsilon_{k_{F}}^{2}$ (Supplemental Material). Indeed,
Shubnikov--de Haas experiments indicate the presence of very heavy
quasiparticles as the field is tilted towards the $z$-axis \cite{Ter02.1}.
Assuming again that $\tau_{\bk}$ is isotropic, the velocity (or effective mass)
anisotropy then leads to corresponding mean free path and Hall conductivity
anisotropies.

Thus, the curvature and hybridization induced mass anisotropies can both give
rise to the same Hall anisotropies. We now demonstrate that the anisotropy of
the Nernst conductivity $\alpha_{xy}$ distinguishes between the two. According
to the Mott formula, $\alpha_{xy}$ is determined by the energy derivative of the
Hall conductivity
\begin{equation}\label{Mott}
\alpha_{xy} = \left. Q_{0}T
\frac{\partial \sigma_{xy}}{\partial E}\right\vert_{E_F}
\end{equation}
with $Q_{0}=\frac{\pi^{2}k_{B}^{2}}{3e}$. Combining this with the Ong formula
(\ref{Ong}) we see that the Nernst conductivity is sensitive to the energy
dependence of the mean free path around the Fermi surface. In the one-band
picture, this anisotropy is entirely determined by the Fermi surface curvature,
and thus the Hall and the Nernst signals share the same anisotropy. By contrast,
in the hybridization picture, the energy dependence of the mean free path is
distinct from the Fermi surface curvature. 

To understand this in more detail, it is useful to consider the normalized ratio
of the Nernst and Hall conductivities (Nernst-Hall ratio)
\begin{equation}\label{NHratio}
D^{*}_{xy} \equiv \left(\frac{\alpha_{xy}}{Q_{0}T} \right)
\frac{1}{\sigma_{xy}} \quad .
\end{equation}
Using the Mott formula (\ref{Mott}) we see that this is the logarithmic energy
derivative of the Hall conductivity
\begin{equation}\label{NHratio}
D^{*}_{xy} = \left. 
\frac{\partial\ln \sigma_{xy}}{\partial E}\right\vert_{E_F}\quad .
\end{equation}
With the Ong formula (\ref{Ong}) this results in
\begin{equation}\label{}
D^{*}_{xy} = \left.
\frac{\partial\ln}{\partial E} \left(\int dk_{z}\cdot
\vec{z}\cdot\oint \vec{l}\times d\vec{l}\right)\right\vert_{E_F} \quad ,
\end{equation}
where the integral over $k_{z}$ is the direction perpendicular to the $xy$
plane. Thus, $D^{*}_{xy}$ is a measure of those regions of the quasiparticle
orbit in which the area swept out by the mean free path is most sensitive to the
chemical potential.

In a one-band picture, changing the Fermi energy $E_{F}$ does not affect the
aspect ratios of the Fermi surface. In a simple relaxation time approximation
the Fermi momenta, velocities, and mean free paths are all proportional to the
square-root of the Fermi energy so that, assuming $\tau_\bk$ to be independent
of energy, the logarithmic derivative of the conductivity is given by 
\begin{equation}\label{NH1band}
D^{*}_{xy} = D^{*}_{yz} = D^{*}_{zx} = \frac{3}{2}\frac{1}{E_{F}} \quad ,
\end{equation}
independent of the direction of measurement. In other words, the anisotropies in
densities of states and mean free path compensate one another in all directions
and thus the Nernst-Hall ratio is isotropic. A strongly anisotropic Hall conductivity, as
observed for CeNiSn, would thus be accompanied by a Nernst conductivity with a
qualitatively similar anisotropy, which is at odds with our Nernst measurements
on CeNiSn.

However, in the hybridized two-band picture, a change in the Fermi energy
produces a very large change in the Fermi momentum of the heavy hybridized band
in the plane perpendicular to the nodal axis (Fig.\,\ref{images} (b) right, see
Supplemental Material for details), but only a small change in Fermi momentum in
the unhybridized direction along the $z$-axis. 

In the Supplemental Material we show that in the relaxation time approximation,
with an energy independent $\tau_\bk$, the ratio of Fermi energy to Kondo
energy enters in the expression for the Hall conductivity anisotropy but
drops out of the corresponding relation for the Nernst conductivity, leading to the Nernst-Hall ratios
\begin{eqnarray}\label{l}
D^{*}_{yz}= D^{*}_{zx}= \frac{1}{E_{F}} \quad \mbox{and} \quad D^{*}_{xy}= \frac{1}{2E_{K}}
\end{eqnarray}
and an essentially isotropic Nernst conductivity.

In experiment, it is the  Nernst coefficient $N_{xy}$ rather than the Nernst
conductivity $\alpha_{xy}$ which is measured. In this experiment $E_z = \nabla_z
T = 0$. Using in addition $\nabla T_y \approx 0$ and $\sigma_{xy}^2 \ll
\sigma_{xx}\sigma_{yy}$ as estimated to be valid to better than 10\% for the
data shown in Fig.\,\ref{ratio}, we obtain
\begin{equation}
\alpha_{yx} \approx \sigma_{yx} N_{xx} + \sigma _{yy} N_{yx} \quad .
\label{Eq2}
\end{equation}
For CeNiSn we find that, at 1\,T, the first term is negligible below 8\,K and thus
\begin{equation}
\alpha_{yx} \approx \sigma_{yy} N_{yx} \quad .
\label{Eq3}
\end{equation}
In addition, the Nernst ratio is close to 1, in particular below 4 K, where it
lies between 0.8 and 1.2. Therefore, the ratio
$\alpha_{zx}/\frac{1}{2}(\alpha_{xy}+\alpha_{yx})$ follows the ratio of the
electrical conductivities at low temperatures (Fig.\,\ref{ratio}) and reaches
thus values of up to 4. Though less isotropic than the Nernst signal itself, the
Nernst conductivities are still much less anisotropic than the Hall
conductivities.

The dichotomy between an isotropic Nernst and a strongly anisotropic Hall
conductivity provides a valuable signature of nodal hybridization in failed
Kondo insulators. One of the unsolved mysteries of Kondo insulators is that all
known true Kondo insulators are cubic, whereas the putative nodal Kondo
semimetals, including CeNiSn, CeRhSb, and CeRu$_{4}$Sn$_{4}$ \cite{Gur13.1}
develop a conducting pseudogap. In the ongoing search for topological Kondo
systems \cite{Dze10.1,Den13.1,Xu14.1s} it is important to understand why the
departure from cubic behavior leads, seemingly innevitably, to semimetallic
behavior. One interesting possibility is that these systems are related to
nodal-line semimetals \cite{Fan15.1}. The Nernst-Hall dichotomy that we have
discovered will provide a useful way to confirm nodal behavior in these systems.

An extreme version of nodal semimetallic behavior has been hypothesized to occur
in the quantum critical semimetal $\beta$-YbAlB$_4$
\cite{Nak08.1,Mat11.1,Ram12.1}. In this system, the presence of a double vortex
in the hybridization node is a tentative explanation of the singular density of
states that develops under ambient pressure conditions. Measurements of the Hall
and Nernst anisotropies may provide a way to confirm the nodal hypothesis in
this system in future work.

In conclusion, we have measured the Nernst coefficient for the Kondo semimetal
CeNiSn, showing that the dichotomy between the severely anisotropic Hall
conductivity and the giant, yet isotropic Nernst signal can be understood in
terms of the nodal hybridization theory of this system. One of the fascinating
open questions is whether the semimetallic behavior of CeNiSn is, in any way,
topologically protected by the crystal symmetries, as in the case of Weyl
semimetals. One of the unexplored and interesting issues of this line of
thought is whether CeNiSn might possess novel surface states, an aspect that has
not been considered in our current analysis. We hope that the Hall-Nernst
dichotomy will provide a new impetus for experimental and theoretical work to
address these open questions.  \vspace{0.2cm}

We gratefully acknowledge financial support from the Austrian Science Fund Grant
I623-N16, the European Research Council Advanced Grant 227378, the U.S.\ Army
Research Office Grant W911NF-14-1-0497 (SP), the National Science Foundation
Grant DMR 1309929 (PC), and the KAKENHI Grant 26400363 (TT). This work was
performed in part at the Aspen Center for Physics, which is supported by
National Science Foundation grant PHY-1066293.

\vspace{0.2cm}

$^{\ast}$Present address: Chinese Academy of Sciences, Institute of Physics,
Beijing National Laboratory for Condensed Matter Physics, Beijing 100190, China


\begin{thebibliography}{10}

\bibitem{Fu07.1}
L. Fu, C.~L. Kane, and E.~J. Mele, {Phys.\ Rev.\ Lett.} {\bf 98},  106803
  (2007).

\bibitem{Moo10.1}
J.~E. Moore, Nature {\bf 464},  194  (2010).

\bibitem{Ali12.1}
J. Alicea, {Rep.\ Prog.\ Phys.} {\bf 75},  076501  (2012).

\bibitem{Wan11.2}
X. Wan, A.~M. Turner, A. Vishwanath, and S.~Y. Savrasov, {Phys.\ Rev.\ B} {\bf
  83},  205101  (2011).

\bibitem{Yan11.4}
K.-Y. Yang, Y.-M. Lu, and Y. Ran, {Phys.\ Rev.\ B} {\bf 84},  075129  (2011).

\bibitem{Dze10.1}
M. Dzero, K. Sun, V. Galitski, and P. Coleman, {Phys.\ Rev.\ Lett.} {\bf 104},
  106408  (2010).

\bibitem{Dez16.1}
M. Dzero, J. Xia, V. Galitski, and P. Coleman, {Annu.\ Rev.\ Condens.\ Matter
  Phys.} {\bf 7},  null  (2016).

\bibitem{Den13.1}
J.~D. Denlinger, J.~W. Allen, J.-S. Kang, K. Sun, J.-W. Kim, J. Shim, B.~I.
  Min, D.-J. Kim, and Z. Fisk, {arXiv: 1312.6637}.

\bibitem{Xu14.1s}
{N.\ Xu et al.}, {Nature Commun.} {\bf 5},  {4566}  (2014).

\bibitem{Ike96.1}
H. Ikeda and K. Miyake, J.\ Phys.\ Soc.\ Jpn. {\bf {65}},  {1769}  ({1996}).

\bibitem{Mor00.1}
J. Moreno and P. Coleman, {Phys.\ Rev.\ Lett.} {\bf 84},  342  (2000).

\bibitem{Nak96.1}
K. Nakamura, Y. Kitaoka, K. Asayama, T. Takabatake, G. Nakamoto, H. Tanaka, and
  H. Fujii, {Phys.\ Rev.\ B} {\bf {53}},  {6385}  ({1996}).

\bibitem{Nak96.2}
{K.~Nakamura, Y.~Kitaoka, K.~Asayama, T.~Takabatake, G.~Nakamoto, and
  H.~Fujii}, Phys.\ Rev.\ B {\bf {54}},  {6062}  ({1996}).

\bibitem{Eki95.1}
{T.~Ekino, T.~Takabatake, H.~Tanaka, and H.~Fujii}, Phys.~Rev.~Lett. {\bf
  {75}},  {4262}  ({1995}).

\bibitem{Dav97.1}
D.~N. Davydov, S. Kambe, A.~G.~M. Jansen, P. Wyder, N. Wilson, G. Lapertot, and
  J. Flouquet, {Phys.\ Rev.\ B} {\bf 55},  R7299  (1997).

\bibitem{Kyo90.1}
M. Kyogaku, Y. Kitaoka, H. Nakamura, K. Asayama, T. Takabatake, F. Teshima, and
  H. Fujii, {J.\ Phys.\ Soc.\ Jpn.} {\bf {59}},  {1728}  ({1990}).

\bibitem{Nak94.1}
{K.~Nakamura, Y.~Kitaoka, K.~Asayama, T.~Takabatake, H.~Tanaka, and H.~Fujii},
  J.\ Phys.\ Soc.\ Jpn. {\bf {63}},  {433}  ({1994}).

\bibitem{Tak96.2}
{T.~Takabatake, G.~Nakamoto, M.~Sera, K.~Kobayashi, H.~Fujii, K~.Maezawa,
  I.~Oguro, and Y.~Matsuda}, J.\ Phys.\ Soc.\ Jpn. {\bf {65, Suppl. B}},  {105}
   ({1996}).

\bibitem{Ina96.1}
Y. Inada, H. Azuma, R. Settai, D. Aoki, Y. \={O}nuki, T. Takabatake, G.
  Nakamoto, H. Fujii, and K. Maezawa, {J.\ Phys.\ Soc.\ Jpn.} {\bf 65},  1158
  (1996).

\bibitem{Tak98.1}
{T.~Takabatake, F.~Iga, T.~Yoshino, Y.~Echizen, K.~Katoh, K.~Kobayashi,
  M.~Higa, N.~Shimizu, Y.~Bando, G.~Nakamoto, H.~Fujii, K.~Izawa, T.~Suzuki,
  T.~Fujita, M.~Sera, M.~Hiroi, K.~Maezawa, S.~Mock, H.~v.~L\"ohneysen,
  A.~Br\"uckl, K.~Neumaier, and K.~Andres}, J.\ Magn.\ Magn.\ Mater. {\bf
  {177-181}},  {277}  ({1998}).

\bibitem{Ter02.1}
T. Terashima, C. Terakura, S. Uji, H. Aoki, Y. Echizen, and T. Takabatake,
  Phys. Rev. B {\bf 66},  075127  (2002).

\bibitem{Ong91.1}
N.~P. Ong, Phys. Rev. B {\bf 43},  193  (1991).

\bibitem{Nak95.1}
{G.~Nakamoto, T.~Takabatake, H.~Fujii, A.~Minami, K.~Maezawa, I.~Oguro, and
  A.A.~Menovsky}, {J.\ Phys.\ Soc.\ Jpn.} {\bf {64}},  {4834}  ({1995}).

\bibitem{Pou06.1}
A. Pourret, K. Behnia, D. Kikuchi, Y. Aoki, H. Sugawara, and H. Sato, Phys.
  Rev. Lett. {\bf 96},  176402  (2006).

\bibitem{Bel04.1}
R. Bel, H. Jin, K. Behnia, J. Flouquet, and P. Lejay, Phys. Rev. B {\bf 70},
  220501  (2004).

\bibitem{Bel04.2}
R. Bel, K. Behnia, Y. Nakajima, K. Izawa, Y. Matsuda, H. Shishido, R. Settai,
  and Y. \={O}nuki, Phys. Rev. Lett. {\bf 92},  217002  (2004).

\bibitem{Mot80.1}
N.~F. Mott, {J.\ Physique Colloque} {\bf 41},  51  (1980).

\bibitem{Aep92.1}
G. Aeppli and Z. Fisk, {Comments Condens.\ Matter Phys.} {\bf 16},  155
  (1992).

\bibitem{Tsu97.1}
H. Tsunetsugu, M. Sigrist, and K. Ueda, {Rev.\ Mod.\ Phys.} {\bf 69},  809
  (1997).

\bibitem{Ris00.1}
P.~S. Riseborough, {Adv.\ Phys.} {\bf 49},  257  (2000).

\bibitem{Gur13.1}
V. Guritanu, P. Wissgott, T. Weig, H. Winkler, J. Sichelschmidt, M. Scheffler,
  A. Prokofiev, S. Kimura, T. Iizuka, A.~M. Strydom, M. Dressel, F. Steglich,
  K. Held, and S. Paschen, {Phys.\ Rev.\ B} {\bf 87},  115129  (2013).

\bibitem{Fan15.1}
C. Fang, Y. Chen, H.-Y. Kee, and L. Fu, {Phys.\ Rev.\ B} {\bf 92},  081201
  (2015).

\bibitem{Nak08.1}
S. Nakatsuji, K. Kuga, Y. Machida, T. Tayama, T. Sakakibara, Y. Karaki, H.
  Ishimoto, S. Yonezawa, Y. Maeno, E. Pearson, G. Lonzarich, L. Balicas, H.
  Lee, and Z. Fisk, {Nature Phys.} {\bf 4},  603  (2008).

\bibitem{Mat11.1}
Y. Matsumoto, S. Nakatsuji, K. Kuga, Y. Karaki, N. Horie, Y. Shimura, T.
  Sakakibara, A.~H. Nevidomskyy, and P. Coleman, Science {\bf 331},  316
  (2011).

\bibitem{Ram12.1}
A. Ramires, P. Coleman, A. Nevidomskyy, and A. Tsvelik, {Phys.\ Rev.\ Lett.}
  {\bf 109},  176404  (2012).

\end{thebibliography}

\end{document}